\documentclass[ajl]{emulateapj}
\usepackage{graphicx,epsfig,natbib,color}
\usepackage{lscape}
\usepackage{graphicx}
\usepackage{epsfig}
\usepackage{natbib}
\begin{document}
\title{On the Relationship Between a Hot-channel-like Solar Magnetic Flux Rope and its embedded Prominence}
\author{X. Cheng$^{1,2,3}$, M. D. Ding$^{1,3}$, J. Zhang$^{1,4}$, A. K. Srivastava$^{5}$, Y. Guo$^{1,3}$, P. F. Chen$^{1,3}$ \& J. Q. Sun$^{1,3}$}
\affil{$^1$ School of Astronomy and Space Science, Nanjing University, Nanjing 210093, China}\email{xincheng@nju.edu.cn}
\affil{$^2$ Sate Key Laboratory of Space Weather, Chinese Academy of Sciences, Beijing 100190}
\affil{$^3$ Key Laboratory for Modern Astronomy and Astrophysics (Nanjing University), Ministry of Education, Nanjing 210093, China}
\affil{$^4$ School of Physics, Astronomy and Computational Sciences, George Mason University, Fairfax, VA 22030, USA}
\affil{$^5$ Department of Physics, Indian Institute of Technology (Banaras Hindu University), Varanasi-21005, India}

\begin{abstract}
Magnetic flux rope (MFR) is a coherent and helical magnetic field structure that is recently found probably to appear as an elongated hot-channel prior to a solar eruption. In this paper, we investigate the relationship between the hot-channel and associated prominence through analyzing a limb event on 2011 September 12. In the early rise phase, the hot-channel was cospatial with the prominence initially. It then quickly expanded, resulting in a separation of the top of the hot-channel from that of the prominence. Meanwhile, both of them experienced an instantaneous morphology transformation from a $\Lambda$ shape to a reversed-Y shape and the top of these two structures showed an exponential increase in height. These features are a good indication for the occurrence of the kink instability. Moreover, the onset of the kink instability is found to coincide in time with the impulsive enhancement of the flare emission underneath the hot-channel, suggesting that the ideal kink instability likely also plays an important role in triggering the fast flare reconnection besides initiating the impulsive acceleration of the hot-channel and distorting its morphology. We conclude that the hot-channel is most likely the MFR system and the prominence only corresponds to the cool materials that are collected in the bottom of the helical field lines of the MFR against the gravity.
\end{abstract}

Online-only material: animations, color figures

\section{Introduction}
A magnetic flux rope (MFR) is a coherent magnetic structure with magnetic field lines wrapping around its central axis. It has been used as a significant configuration to study the initiation mechanisms of solar energetic phenomena including flares, prominences (or filaments when seen on the solar disk), and coronal mass ejections (CMEs) \citep[e.g.,][]{fan04,torok05,kliem06,aulanier10,olmedo10,leake13,nishida13}. Based on whether magnetic reconnection is involved or not in destabilization process, present models can be largely grouped into two categories. One is reconnection type including inner tether-cutting \citep{moore01} and top/lateral breakout reconnections \citep{antiochos99,chen00}. The other type is the ideal magnetohydrodynamic instabilities of the MFR including torus instability \citep{kliem06} and/or kink instability \citep{hood81}. The torus instability occurs if the restoring force of the MFR caused by the background field decreases faster than the outward-directed Lorenz self-force as the MFR expands \citep{kliem06,olmedo10}. The kink instability refers to the helical instability of the MFR, which takes place if the average twist number of the MFR exceeds a threshold \citep{torok04,srivastava10}. 

Because of the theoretical importance of the MFR, researchers concern the question of whether the MFR exists prior to the eruption. To the present, indirect evidence that supports pre-existence of the MFR has been uncovered such as forward or reversed S-shaped sigmoids \citep{rust96,canfield99,tripathi09} and dark cavities \citep{low95_apj,gibson04,dove11,bak-steslicka13}. Filaments are also thought to be the evidence of the pre-existence of the MFR because they often correspond well to the dips of the helical lines in extrapolated nonlinear force-free field configurations \citep[e.g.,][]{mackay10,guo10_filament,suyingna11}. Recently, utilizing the Atmospheric Imaging Assembly \citep[AIA;][]{lemen12} telescope on board the \textit{Solar Dynamics Observatory} (\textit{SDO}), \citet{zhang12} and \citet{cheng13_driver} discovered another important structure: a coherently elongated and S-shaped hot-channel. It appears above the neutral line of active region tens of minutes before the eruption and is only visible in the AIA 131 {\AA} and 94 {\AA} passbands, showing high temperatures of $\ge$8 MK. As the hot-channel ascends, its morphology quickly transforms to a loop-like shape. However, during the transformation process, the two footpoints remain fixed in the photosphere. With the channel expanding and rising up, a CME is quickly built up in the very low corona \citep[also see;][]{liur10,cheng13_double,patsourakos13,lileping13}. These results strongly suggest that the hot-channel is most likely to be the MFR. In this Letter, we further investigate the relationship between the hot-channel and associated prominence through a detailed analysis of a limb event on 2011 September 12. In Section \ref{ss2}, we show data reduction and results, followed by the summary and discussions in Section \ref{ss3}.

\begin{figure*}
\center {\includegraphics[width=16cm]{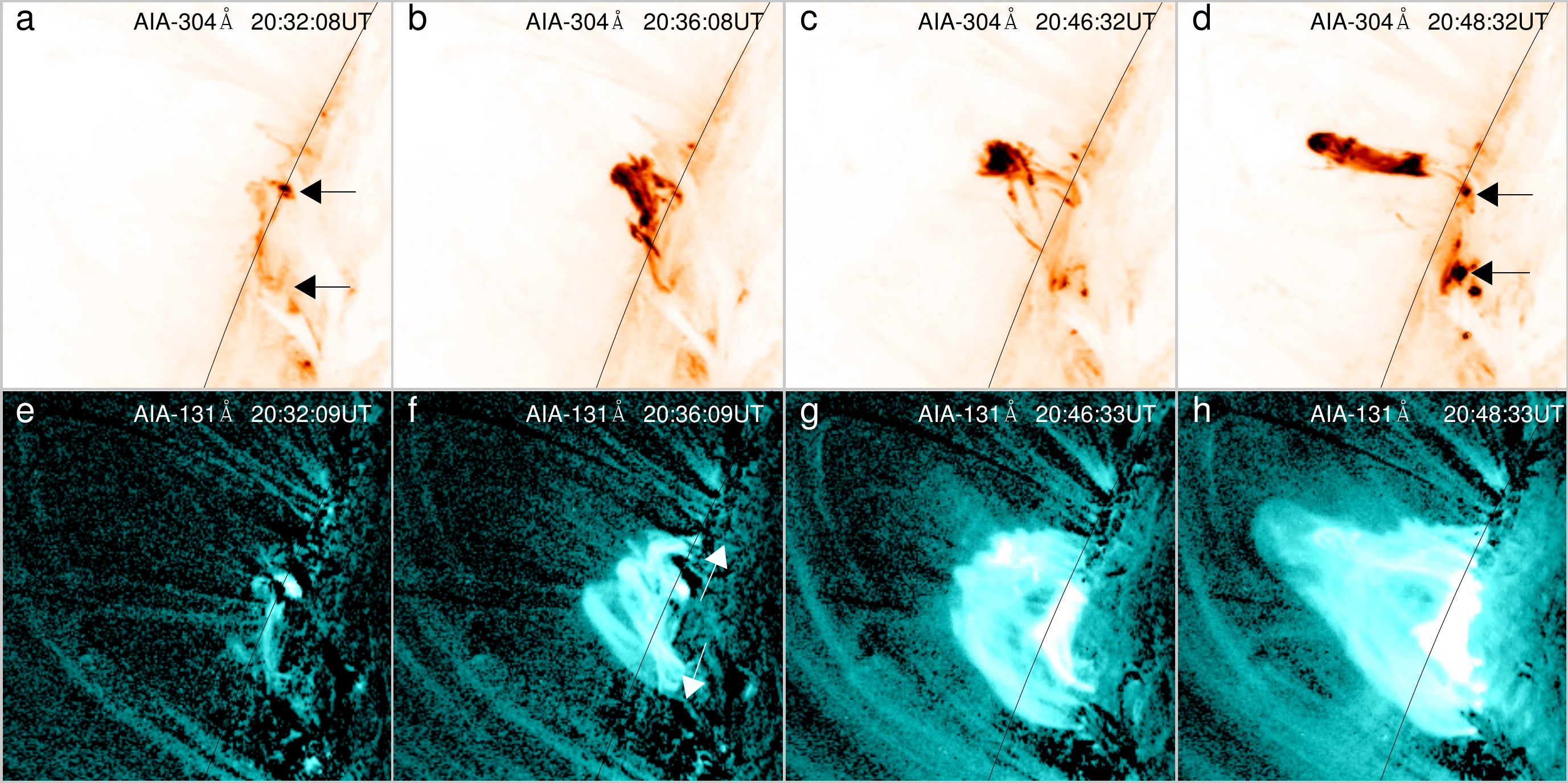}}
\caption{a--d: SDO/AIA 304 {\AA} ($\sim$0.05 MK) (negative intensity) images displaying the heating and eruption of the prominence on 2011 September 12. Two black arrows in panel a and e indicate the two footpoints of the prominence. e--h: SDO/AIA 131 {\AA} ($\sim$0.4 MK and 11.0 MK) base-difference images, with the base image at 20:20 UT, showing the heating and eruption of the hot-channel. Two white arrows in panel f point out the extension of the hot-channel footpoints in the early phase. The black line in each panel corresponds to the solar limb.}
(Animation this figure are available in the online journal.)
\label{f_fr}
\end{figure*}

\section{Observations and Results}\label{ss2}
\subsection{Heating and Eruption of the Prominence}
On 2011 September 12, a prominence erupted above the east limb of the Sun. Its early evolution was well captured by the \textit{SDO}/AIA thanks to its high spatial resolution (1.2\arcsec), high temporal cadence (12 s), and multi-temperature \citep[0.06--20 MK;][]{odwyer10} ability. \citet{tripathi13} studied this prominence but concentrated on its partial eruption. Here, we focus on the relationship between the prominence and associated hot-channel-like MFR. The early activation was clearly seen from the AIA 304 {\AA} images (Figure \ref{f_fr}a and \ref{f_fr}b). Initially, the prominence lay low above the solar surface with the two footpoints anchored in the chromosphere (Figure \ref{f_fr}a). Probably due to the reconnection evidenced by the EUV brightenings along the prominence and the slight enhancement of the \textit{GOES} soft X-ray (SXR) 1--8 {\AA} flux \citep[also see][]{tripathi13}, the prominence started to rise slowly after $\sim$20:30 UT (Figure \ref{f_fr}b). At $\sim$20:46 UT, the prominence suddenly exhibited an impulsive acceleration. The morphology evolved instantaneously from a $\Lambda$ shape to a reversed-Y shape (Figure \ref{f_fr}c and \ref{f_fr}d), inferring the occurrence of the kink instability \citep[e.g.,][]{torok05}. However, probably due to that some materials drained down to the chromosphere, the left part of the reversed-Y shape became invisible since $\sim$20:48 UT. Moreover, at $\sim$20:46 UT, the brightness at the projected crossing part of the two legs of the prominence increased to the maximum in all AIA EUV passbands (Figure \ref{f_fr}c). It implies that magnetic reconnection took place there. Subsequently, the EUV brightenings were also enhanced at the two footpoints of the prominence (Figure \ref{f_fr}d), indicating that the reconnection also heats the chromosphere. 

\subsection{Heating and Eruption of the Hot-channel-like MFR} 
We find that the erupted prominence was closely associated with an elongated hot-channel-like structure. In the first serval minutes, some diffuse threads, similar to the prominence, were visible in the AIA 131 {\AA} passbands (Figure \ref{f_fr}e). With time elapsing, probably due to the reconnection heating, more and more threads revealed themselves. At 20:36 UT (Figure \ref{f_fr}f), all of the threads seemed to be converged together. At 20:46 UT, the whole system formed a well-shaped and coherent channel-like structure (Figure \ref{f_fr}g). As revealed in the previous studies \citep{zhang12,cheng13_driver,patsourakos13}, the structure can only be seen in the AIA high temperature passbands, i.e., 131 {\AA} and 94 {\AA}, but not in the other cooler wavelengths. It shows that the channel must have a temperature of $\ge$8 MK. 

\begin{figure*}
\center {\includegraphics[width=10cm]{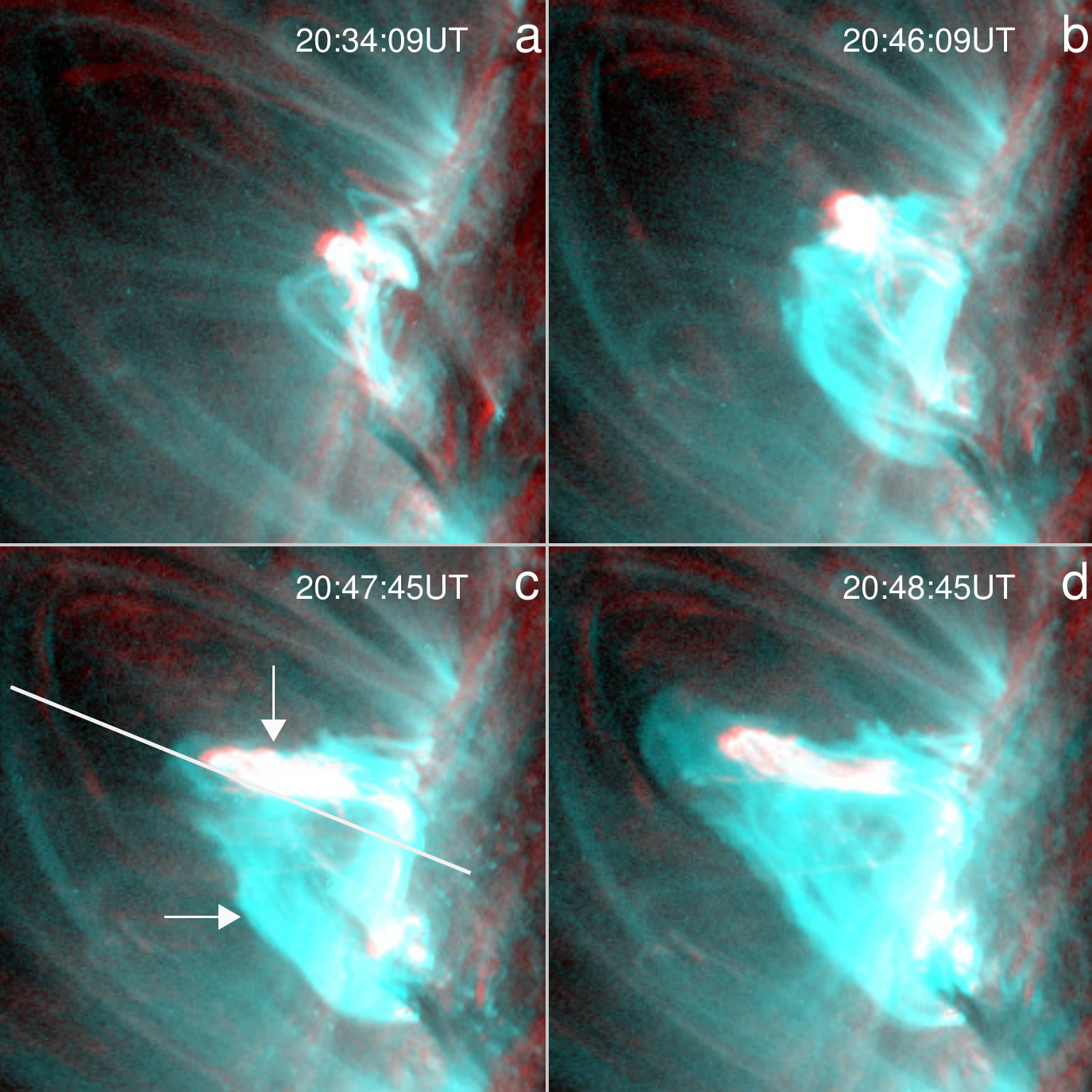}\vspace{-0.03\textwidth}}
\center {\includegraphics[width=11.9cm]{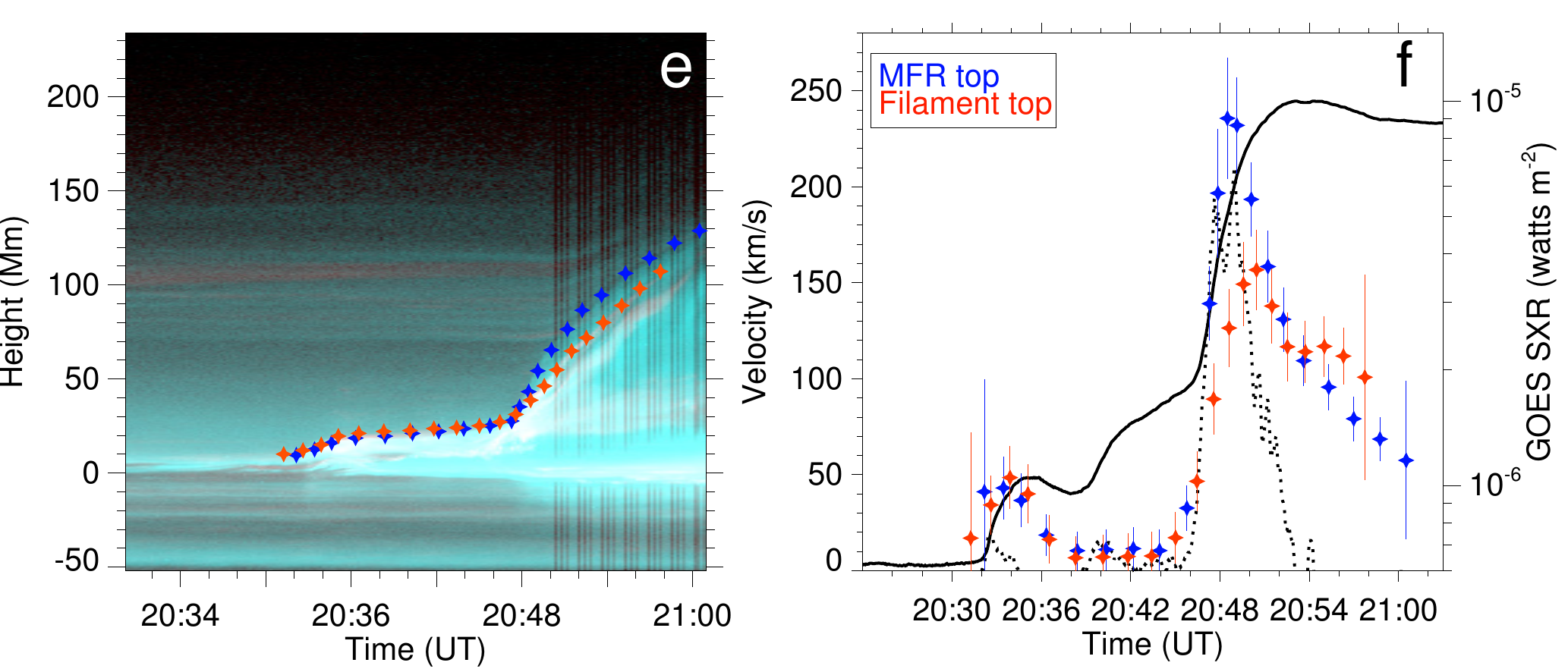}\vspace{-0.0\textwidth}}
\caption{a--d: Composite of the AIA 131~{\AA} ($\sim$0.4 MK and 11.0 MK) and 304~{\AA} ($\sim$0.05 MK) images showing the spatial relationship between the hot-channel (blue) and associated prominence (light red) before and during the eruption. e: Stack plot of the AIA 131 {\AA} and 304 {\AA} composite intensity along the slice as shown by the white oblique line in panel c. Blue and red stars indicate the height measurements of the hot-channel and prominence above the solar surface, respectively. f: Velocity evolution of the hot-channel (blue) and prominence (red). The solid and dotted curves denote the GOES 1--8~{\AA} SXR flux of the associated flare and its time derivative, respectively.}
(An animation this figure is available in the online journal.)
\label{f_kin}
\end{figure*}

An interesting finding is that initially the hot-channel was almost co-aligned with the prominence in space but later on the top of the hot-channel separated from that of the prominence (Figure \ref{f_kin}a--\ref{f_kin}c). In particular, at $\sim$20:48 UT, the prominence was only cospatial with the bottom of the right part of the hot-channel (Figure \ref{f_kin}d). The results can be interpreted as a general scenario that the hot-channel is likely the MFR and the prominence is only the collection of the cool materials at the bottom part of the MFR; the eruption of the prominence was essentially followed by that of the MFR. Moreover, the hot-channel also displayed the similar morphology transformation like the prominence. From $\sim$20:46 UT, the top of the hot-channel started to arch upward. In the period of 20:46--20:48 UT, the arching first made the hot-channel take on the $\Lambda$ shape (Figure \ref{f_kin}c), which then quickly evolved into the reversed-Y shape (Figure \ref{f_fr}h and Figure \ref{f_kin}d).

As the morphology of the hot-channel transited from the $\Lambda$ to reversed-Y shape, the flare-related reconnection started to dominate the whole heating process. We calculate the differential emission measure (DEM) of the erupted structure through the SolarSoft routine ``xrt{\_}dem{\_}iterative2.pro" \citep{cheng12_dem}. With the DEM results, we then construct the two-dimensional maps of emission measure (EM) of the plasma in different temperature intervals ($\Delta T$) through the formula EM($T$)=$\int_{T-\Delta T}^{T}{\rm DEM}(T^\prime)$d$T^\prime$. 

\begin{figure*}
\center {\includegraphics[width=14cm]{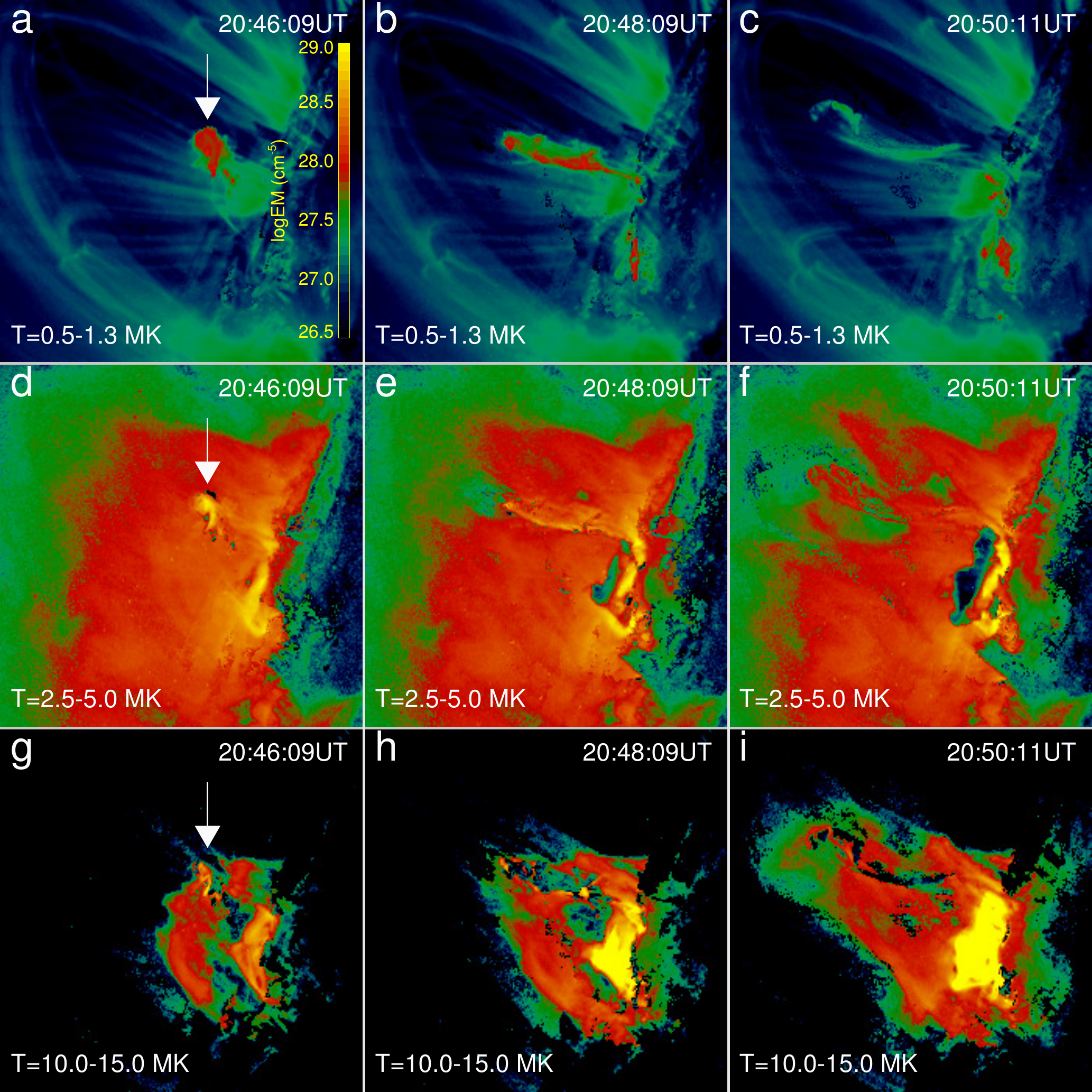}\vspace{-0.0\textwidth}}
\caption{Two-dimensional EM maps of the hot-channel and associated prominence in different temperature intervals. The white arrows denote that the cross part of the prominence has the biggest EM value.}
(An animation this figure is available in the online journal.)
\label{f_dem}
\end{figure*}

Figure \ref{f_dem} shows the EM structures of the eruption in three temperature intervals. It can be seen that, at $\sim$20:46 UT, the plasma in the space occupied by the prominence showed emissions in all temperatures (0.5--5 MK) while the surrounding hot-channel only exhibited emission from the hot plasma (10--15 MK). It reveals that the hot-channel has been heated up to the high temperature at that time. We suggest that magnetic reconnection with a slow rate probably occurs inside or around the hot-channel to heat and build up the channel before $\sim$20:46 UT; however, this reconnection is too weak to generate nonthermal particles, thus being different from the fast flare reconnection in the later phase \citep[also see][]{aulanier10,cheng13_double,guo13_qmap}. After 20:46 UT, the hot-channel still had a temperature of $>$8 MK and the prominence had a temperature of $<$5 MK; however, the EM of both of the hot-channel and prominence tended to decrease, mainly due to expansion. On the other hand, the EM of the flare region underneath the hot-channel was quickly enhanced. In particular, for the hot plasma, the EM increased from $\sim$10$^{28}$ cm$^{-5}$ at 20:47 UT to $\sim$10$^{29}$ cm$^{-5}$ at 20:50 UT in the low-lying flare region. It indicates that the morphology transformation of the hot-channel is also associated with the triggering of the fast flare reconnection that may further help to heat and build up the hot-channel.

\subsection{Kinematical Relationship Between the Hot-channel-like MFR and Prominence}
In this section, we study the kinematics of the hot-channel-like MFR and the prominence in detail. We take a slice along the direction of the eruption (Figure \ref{f_kin}c). The time sequence of the slice makes up a stack plot. Using the stack plot, we measure the heights of the hot-channel and associated prominence. Applying the first order piecewise numerical derivative to the height-time data, we derive the velocities of the hot-channel and prominence. The uncertainties in the velocities arise from the uncertainties in the height measurements, which are estimated to be 4 pixel sizes ($\sim$1.7 Mm). 

Figure \ref{f_kin}e and \ref{f_kin}f display the height and velocity profiles of the hot-channel and prominence. One can find that the hot-channel experienced two distinct phases: a slow rise phase of twenty minutes and an impulsive acceleration phase of only $\sim$5 minutes. During the first several minutes, the height of the hot-channel increased from $\sim$10 Mm at 20:30 UT to $\sim$20 Mm at 20:36 UT, resulting in an average velocity of $\sim$30 km s$^{-1}$. The early rise process well corresponds to the brightening along the hot-channel and prominence, confirming that their activation was most likely the result of the slow reconnection. In the period of $\sim$10 minutes after 20:36 UT, the rise of the hot-channel tended to slow down; the height only increased to 25 Mm at $\sim$20:46 UT, corresponding to an average velocity of $\sim$10 km s$^{-1}$ (Figure \ref{f_kin}f). 

\begin{figure*}
\center {\includegraphics[width=13cm]{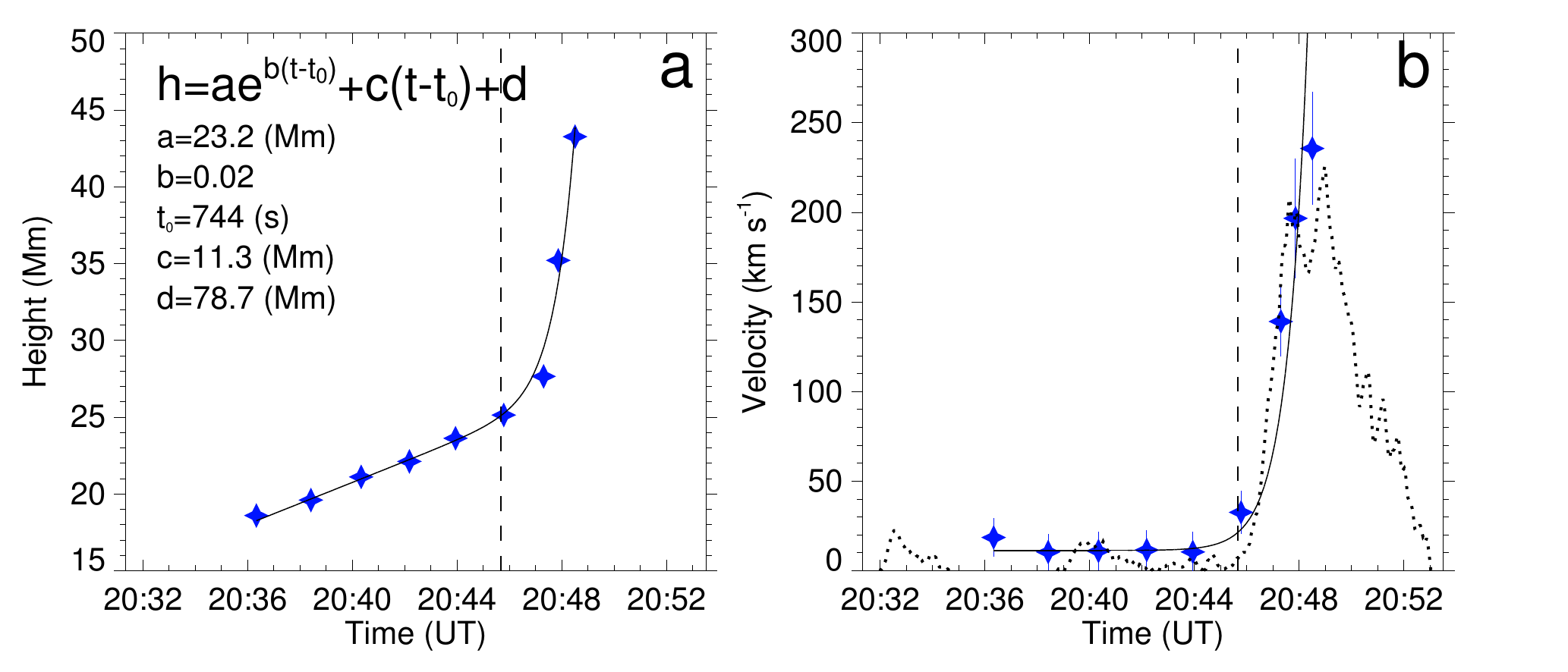}\vspace{-0.0\textwidth}}
\center {\includegraphics[width=13cm]{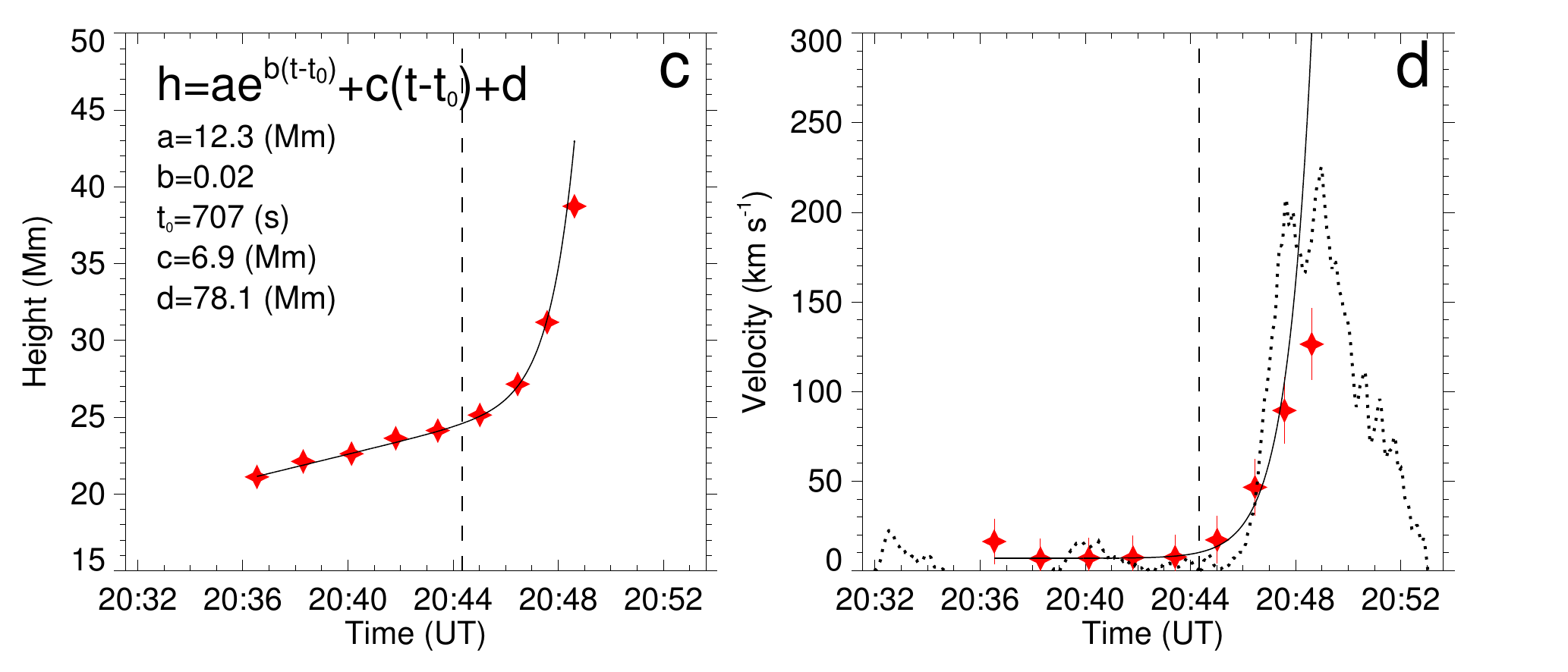}\vspace{-0.0\textwidth}}
\caption{a and b: Best fitting of the height and velocity of the hot-channel, as shown by the solid lines. The fitting function and resulting parameters are displayed at the top left corner. c and d: Same as a and b but for the prominence. The dotted line shows the \textit{GOES} derivative and the two vertical dashed lines show the onset time of the impulsive acceleration.}
\label{f_kin2}
\end{figure*}

An important result is that the height evolution of the hot-channel had an apparent jump (Figure \ref{f_kin}e). Correspondingly, the velocity impulsively increased around the jump, e.g., from $\sim$10 km s$^{-1}$ at $\sim$20:46 UT to $\sim$250 km s$^{-1}$ three minutes later at $\sim$20:49 UT (Figure \ref{f_kin2}b). The average acceleration in this period is estimated to be $\sim$1300 m s$^{-2}$, which is much larger than the average acceleration (330 m s$^{-2}$) of the impulsive CMEs \citep[e.g.,][]{zhang06}. After $\sim$20:49 UT, the velocity started to decrease gradually and became to be $\sim$50 km s$^{-1}$ at $\sim$21:00 UT with an average deceleration of $\sim$300 m s$^{-2}$. This deceleration led to a failed eruption, as evident from the lack of propagating CME in the white-light coronagraph images. In order to exactly estimate the onset time of the impulsive acceleration, we use a function consisting of both a linear and an exponential component to fit the height-time measurements of the hot-channel from 20:36 UT to 20:48 UT. The details of the technique can be found in \citet{cheng13_double}. From Figure \ref{f_kin2}a and \ref{f_kin2}b, one can see that the height of the hot-channel is well described by the combination of the linear and exponential functions. The exponential component is believed to be a fundamental feature of the MFR eruption trajectory driven by the kink instability \citep[e.g.,][]{torok05,schrijver08_filament}. Assuming that the hot-channel is impulsively accelerated at the time when the velocity of the exponential component is equal to that of the linear component, the impulsive acceleration onset is found to be at 20:45:40 UT with an uncertainty of 1.7 min, thus almost perfectly coincident with the sudden transformation of the hot-channel morphology from the $\Lambda$ shape to reversed-Y shape. These results indicate that the hot-channel most likely underwent the kink instability, thus triggering and driving the impulsive acceleration in a very short period, and meanwhile, distorting the axis of the hot-channel, revealing a transformation of the morphology.

\begin{figure*}
\vspace{-0.3\textwidth}
\center {\includegraphics[width=15cm]{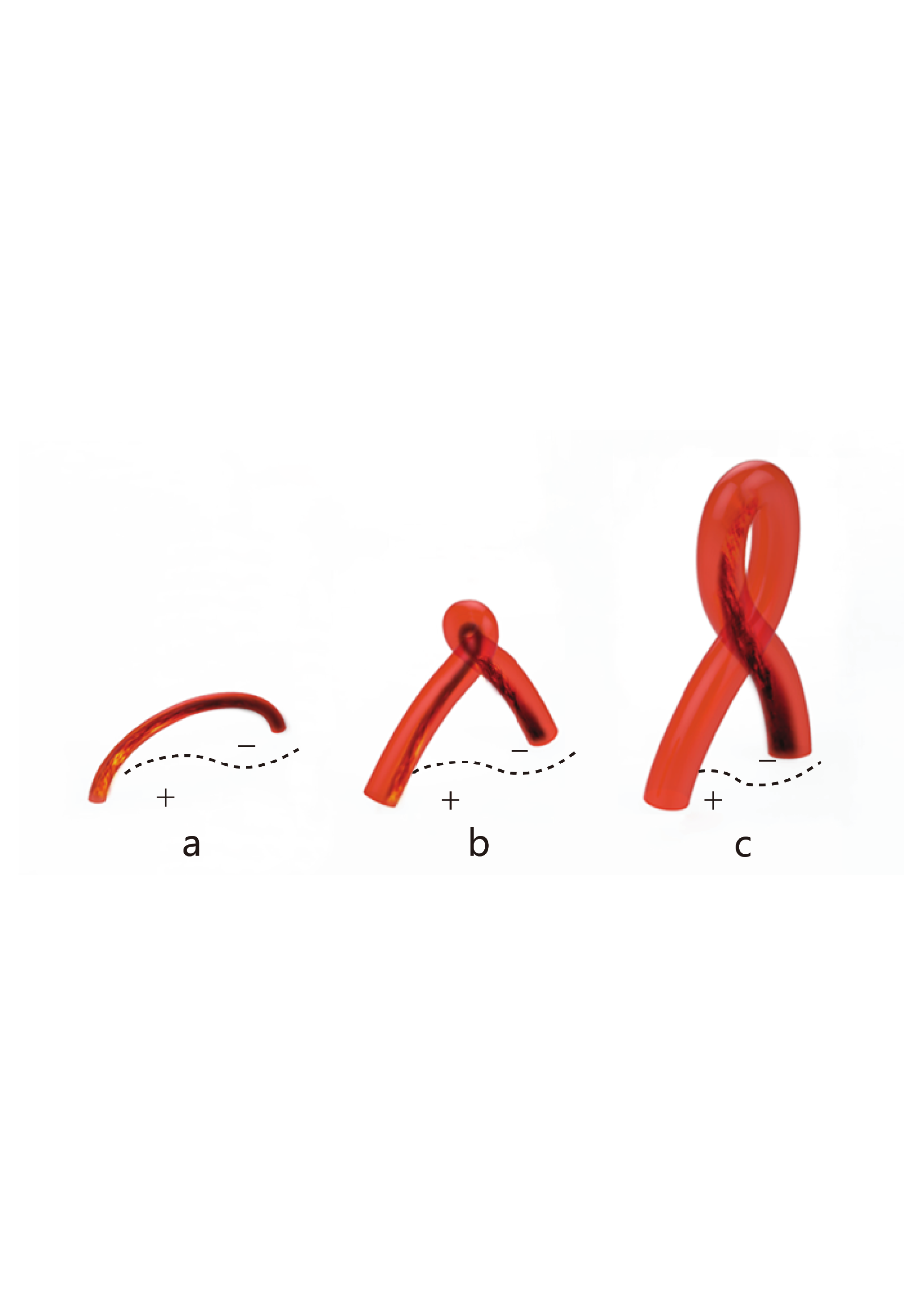}\vspace{-0.38\textwidth}}
\caption{Schematic drawing of the relationship between the hot-channel-like MFR (red tubes) and embedded prominence (dark materials) during the early evolution. a: Pre-eruption structure displaying the cospatiality of the MFR and prominence. b: Separation of the MFR top from the prominence. c: Morphology transformation of the MFR from an $\Lambda$ shape to a reversed-Y shape as a result of kink instability. The dashed curve in each panel shows the polarity inverse line.}
\label{f_cartoon}
\end{figure*}

As for the prominence, the evolution of the height and velocity had exactly the same trends as the hot-channel with, however, some difference in magnitude. In the slow rise phase, the prominence kept the same height and linear velocity as the hot-channel. With the impulsive acceleration commencing, the height of the prominence also exponentially increased (Figure \ref{f_kin2}c). Using the same technique, we are able to fit the height variation of the prominence very well with the combination of the linear and exponential functions. The onset of the prominence impulsive acceleration is determined to be at 20:44:20 UT with an uncertainty of 5.0 min (Figure \ref{f_kin2}d), almost coinciding with that of the hot-channel. On the other hand, we find that the height and velocity of the prominence increased more slowly than that of the hot-channel shortly after the beginning of the impulsive acceleration. These results suggest that the prominence and hot-channel share the same MFR system during the whole eruption course. Most likely, the hot-channel is the MFR and the prominence corresponds to the dips of the helical field lines of the MFR. Prior to and during the slow rise phase, the MFR remains small and compact, thus almost co-spatial with the prominence (Figure \ref{f_kin}a); the distance between the MFR top and prominence is too small to be recognizable. In the acceleration phase, as a result of the fast reconnection, represented by the peak of the time derivative of the \textit{GOES} SXR flux (Figure \ref{f_kin}f), the newly formed high temperature poloidal flux quickly envelops the MFR, resulting in the heating of the MFR and the separation between the tops of the MFR and the prominence; thus the upper part of the MFR is only visible in the AIA high temperature passbands (131 {\AA} and 94 {\AA}), while the lower part is seen in all AIA passbands because this part consists of both the cool core and the hot shell of the MFR (Figure \ref{f_kin}c and \ref{f_kin}d).

Moreover, the onset of the hot-channel and prominence impulsive acceleration was almost coincident with the rapid enhancement of the flare emission (Figure \ref{f_kin2}b and \ref{f_kin2}d). Through inspecting the AIA images, we find that the strongest brightening appeared at the crossing part of the two legs of the hot-channel and prominence at the onset time ($\sim$20:46 UT; Figure \ref{f_fr}c and the left column of Figure \ref{f_dem}). With the hot-channel and prominence ascending, the brightening underneath them was also rapidly increased, showing that the fast reconnection started, which rapidly increased the flare emission and formed the flare loops (Figure \ref{f_fr}g and \ref{f_fr}h; Figure \ref{f_dem}h and \ref{f_dem}i). The transition of the brightening from the crossing part to underneath the hot-channel and prominence implies that the kink instability may also have a role in causing the fast flare reconnection.

\section{Summary and Discussion}\label{ss3}
In this Letter, we investigate the relationship between the hot-channel  and associated prominence. The cospatiality of the prominence with the hot-channel in the early phase (Figure \ref{f_cartoon}a) and a following separation of the top of the hot-channel from that of the prominence in the later phase (Figure \ref{f_cartoon}b) strongly support our previous conjecture that the hot-channel is likely an MFR whose lower part, i.e., the dipped part of the helical magnetic lines, corresponds to the prominence. Using the high cadence AIA data, we find that the evolution of both of them experienced two phases: a slow rise phase and an impulsive acceleration phase. The evolution near the transition from the slow rise to the impulsive acceleration phase can be well described by a combination of the linear and exponential functions. Moreover, the kinematic transition in time coincided with the quick morphological transformation from the $\Lambda$ shape to reversed-Y shape (Figure \ref{f_cartoon}c). These results indicate that the hot-channel most likely underwent the kink instability, thus triggering the impulsive acceleration of the MFR and the fast reconnection producing the flare.

It has been recognized that the impulsive acceleration of the MFR might be triggered by the torus instability \citep{kliem06,fan07,aulanier10,olmedo10,savcheva12b,cheng13_double,cheng14_tracking,zuccarello14}. To tentatively study this possibility, we calculate the three-dimensional magnetic field structure using the magnetic data on 2011 September 17 provided by the Helioseismic and Magnetic Imager \citep{schou12}. We find that the decay index of the background field at the onset heights of the hot-channel and prominence impulsive acceleration ($\sim$25.2$\pm$4.3 Mm and $\sim$25.2$\pm$6.3 Mm) is only $\sim$1.1, which is smaller than the threshold of the torus instability \citep[1.5;][]{kliem06}; thus, the torus instability is unlikely the cause of eruption in this event.

The kink instability of the hot-channel requires a strong twist, thus reinforcing its physical nature as an MFR. Theoretically, the kink instability occurs when the twist number of the MFR is larger than the critical value of 1.5 (3.0$\pi$ in twist angle) for an arched MFR \citep{fan03,torok04}, or at least the threshold of 1.25 for a line-tied cylindrical MFR \citep{hood79,hood81}. This can be considered as another piece of evidence for the existence of the MFR as the hot-channel besides being visually indentified as a bundle of helical threads \citep[e.g.,][]{cheng14_tracking}. Finally, it is worth mentioning that although \citet{tripathi13} studied the same event, they concentrated on the partial eruption of the prominence. Here, we pay much attention to the relationship between the hot-channel and the prominence and conclude that the hot-channel and prominence are two components of the same MFR system that simultaneously rise, accelerate, and deform, subject to the kink instability.

\acknowledgements 
We are grateful to the referee for the valuable comments that have significantly improve the manuscript. SDO is a mission of NASAÕs Living With a Star Program. X.C., M.D.D., Y.G., P.F.C, and J.Q.S are supported by NSFC under grants 11303016, 11373023, 11203014, 11025314, and NKBRSF under grants 2011CB811402 and 2014CB744203. X.C. is also supported by the specialized research fund for state key laboratories. J.Z. is supported by NSF grant ATM-0748003, AGS-1156120, and AGS-1249270. A.K.S thanks P. F. C for the grant to visit Nanjing University where he pursues jointly on various research projects.

%\bibliographystyle{apj}
%\bibliography{reference}

\end{document}